\documentclass[twocolumn,pra,aps,superscriptaddress,nofootinbib]{revtex4-1} 
\usepackage[english]{babel}
\usepackage[utf8]{inputenc} 

\def\N{\mathbb{N}}

\def\Z{\mathbb{Z}}

\usepackage{graphicx}
\usepackage{amsmath,amsthm}
\usepackage{tikz}
\usepackage{amsfonts}
\usepackage{amssymb}
\usepackage{multirow}
\usepackage{mathrsfs}
\usepackage{makeidx}
\usepackage{amsmath}
\usepackage{hyperref}
\usepackage{placeins}

\def\({\left(}
\def\){\right)}
\def\<{\left\langle}
\def\>{\right\rangle}
\def\beq{\begin{equation}}
\def\eeq{\end{equation}}
\def\pl{\partial}

\begin{document}

\title{Shape effects in the fluctuations of random isochrones on
  a square lattice}

\author{Iván Álvarez Domenech}
\affiliation{Dto.\ Física Matemática y de Fluidos, Universidad Nacional
  de Educación a Distancia (UNED), Madrid (Spain)}

\author{Javier Rodríguez-Laguna}
\affiliation{Dto.\ Física Fundamental, Universidad Nacional de
  Educación a Distancia (UNED), Madrid (Spain)}

\author{Rodolfo Cuerno}
\affiliation{Dto.\ Matemáticas \& GISC, Universidad Carlos III de Madrid,
  Leganés (Spain)}

\author{Pedro Córdoba-Torres}
\affiliation{Dto.\ Física Matemática y de Fluidos, Universidad Nacional
  de Educación a Distancia (UNED), Madrid (Spain)}

\author{Silvia N.\ Santalla}
\affiliation{Dto.\ Física \& GISC, Universidad Carlos III de Madrid,
  Leganés (Spain)}

\date{November 1, 2023}

\begin{abstract}
  We consider the isochrone curves in first-passage percolation on a
  2D square lattice, i.e.\ the boundary of the set of points which can
  be reached in less than a given time from a certain origin. The
  occurrence of an instantaneous average shape is described in terms
  of its Fourier components, highlighting a crossover between a
  diamond and a circular geometry as the noise level is increased.
  Generally, these isochrones can be understood as fluctuating
  interfaces with an inhomogeneous local width which reveals the
  underlying lattice structure. We show that once these
  inhomogeneities have been taken into account, the fluctuations fall
  into the Kardar-Parisi-Zhang (KPZ) universality class with very good
  accuracy, where they reproduce the Family-Vicsek Ansatz with the
  expected exponents and the Tracy-Widom histogram for the local
  radial fluctuations.
\end{abstract}

\maketitle

\section{Introduction}

Random curves have attracted attention in many fields of science
\cite{Adler}, such as physics, mathematics, and biology. For instance,
they appear in fields such as polymer physics \cite{Halpin_95},
quantum gravity \cite{Ambjorn_97}, or the characterization of
biophysical objects such as membranes and cells \cite{Nelson}. Let us
focus on {\em isochrone curves} within a random two-dimensional
manifold, i.e.\ the boundaries of balls with different radii, when the
metric is flat on average and presents only short-range correlations.
It was shown in a previous work \cite{Santalla_15,Santalla_17} that in
the continuum these isochrones present a fractal behavior described by
the celebrated Kardar-Parisi-Zhang (KPZ) universality class, which
accounts for the fluctuation statistics of many growing interfaces
\cite{Kardar_86,Kardar_87,Barabasi}. The average roughness of the
isochrone, $W(t)$, defined as the root-mean-square deviation of the
ball radii at time $t$, grows as $W(t)\sim t^{\beta}$, while the
correlation length $\xi$ grows as $\xi\sim t^{1/z}$, where $\beta=1/3$
and $z=3/2$ are respectively the growth exponent and the dynamic
exponent for the 1+1 dimensional (1+1D) KPZ universality class.
Moreover, for long times, the local roughness at a length-scale $l$
behaves as $w(l) \sim l^\alpha$, where $\alpha$ is called the
roughness exponent, which is related to the other two through the
Family-Vicsek relation $\alpha=\beta z$, and takes the value $1/2$ for
the KPZ class.

The discrete analogue of this problem is known as first-passage
percolation (FPP) \cite{Hammersley_65}, which was originally proposed
as a model of fluid flow through random media. The FPP model has
received substantial attention within probability theory, giving rise
to important results such as the sub-additive ergodic theorem
\cite{Kingman}, which has a remarkable relevance for classical
problems such as the Ulam-Hammersley problem, and contributed to
develop the field of integrable probability \cite{Romik}. Moreover,
integrable probability was in turn instrumental to characterize the
one-point and two-point fluctuations within the KPZ universality
class, specifically the emergence of the Tracy-Widom (TW)
distributions \cite{Praehofer_02,Takeuchi_11,Corwin_13}, which were
originally defined as the probability distributions for extreme
eigenvalues in random matrix ensembles \cite{Mehta}. The FPP model has
been thoroughly studied numerically in our previous work
\cite{Cordoba_18,Villarrubia_20}, both in the weak and strong disorder
regimes, confirming the predictions regarding KPZ scaling in the
asymptotic regime for the geodesics. Yet, a similar characterization
of the statistical properties of the isochrones has not been reported
so far, which is the task undertaken in this article. The main
difficulty to this end is the fact that the average shape of the
isochrones is not circular in general, due to the anisotropy of the
lattice. In fact, the existence and characterization of such average
shapes in the long run constitute a relevant area of mathematical
research, which has led to the celebrated {\em shape theorem}
\cite{Cox_81,Damron18}.

More generally, many other contexts for the growth of planar clusters
present analogous complexities, in the sense that non-trivial
interface fluctuations occur around well-defined {\em macroscopic}
shapes. Examples can be found, for instance, in epitaxial growth of
thin solid films in the submonolayer regime \cite{Jensen.99} or in the
spreading of precursor layers of wetting fluids
\cite{Misbah.10,Bonn.09,Marcos.22}. Both of these systems inherently
host strong interface fluctuations due to the small typical scales
which are involved. And also in both cases, being able to subtract
characteristic shapes from front fluctuations can prove significant to
correctly identify the universality class (if appropriate) of the
latter.

This article is organized as follows. We start with a description of
the basic properties of the FPP model in Sec. \ref{sec:model}. In Sec.
\ref{sec:shape} we characterize the instantaneous average shape of the
isochrones for different noise levels using their Fourier components.
The growth and dynamic exponents, $\beta$ and $z$, are determined in
Sec. \ref{sec:timebehavior}, and the histogram of the radial
fluctuations is obtained in Sec. \ref{sec:fluct}. In both cases, the
lattice anisotropy masks the expected KPZ behavior unless the
statistical data are angularly resolved. The roughness exponent,
$\alpha$, presents an additional numerical challenge, because any
uncertainty in the instantaneous average shape may interfere with its
measurement. In Sec. \ref{sec:morphology} we address this issue, and
we find that the aforementioned uncertainty manifests itself as an
intrinsic roughness. The article concludes in Sec.
\ref{sec:conclusions} with a discussion of our main conclusions and
suggestions for further work.

%%%%%%%%%%%%%%%%%%%%%%%%%%%%%%%%%%%%%%%%%%%%%%%%%%%%%%%%%%%%%%%%%%%%%%%

\section{Model}
\label{sec:model}

Let us consider the integer latice $\Z^2$, with edge set $E$. We can
associate a random variable $\tau_e>0$ to each edge $e\in E$, which
we will call its {\em passage time} or {\em link-time}. The variables
$\{\tau_e\}_{e\in E}$ are assumed to be independent, identically
distributed (i.i.d.) with distribution function $F(\tau)$, such that
$F(0)=0$, i.e.\ we assume that $\tau_e$ is strictly positive with
probability one. The associated density function will be denoted by
$f(\tau)$, and $\mu$ and $s$ will denote respectively its mean and
deviation, which we will assume to be finite.

A finite path is defined as a sequence of edges, $e_1,e_2,\ldots,e_n$,
such that $e_i$ and $e_{i+1}$ share exactly one endpoint. For each
path $\gamma=\{e_1,\ldots,e_n\}$ we can define its passage time
$T(\gamma)$ as $T(\gamma)=\sum_i \tau_i$, where the sum runs over all
edges in $\gamma$. Finally, given two different nodes,
$\mathbf{x},\mathbf{y}\in\Z^2$, we can define the passage time between
them, $T(\mathbf{x},\mathbf{y})$, as the minimum passage time over all
paths joining $\mathbf{x}$ with $\mathbf{y}$, which we will denote by

\begin{equation}
T(\mathbf{x},\mathbf{y})=\min_{\gamma\in \Gamma(\mathbf{x},\mathbf{y})} T(\gamma),
\end{equation}
where $\Gamma(\mathbf{x},\mathbf{y})$ denotes the set of paths joining
these two points. The random function $T(\mathbf{x},\mathbf{y})$ plays
the role of a distance, and the pair $\(\Z^2,T(\cdot,\cdot)\)$ defines
a metric space in which the geodesic between two nodes is given by the
path of minimal arrival time~\cite{Cordoba_18}. The FPP model is
mathematically equivalent to the problem of {\em optimal paths} in
weighted networks \cite{StanleyPRL96,StanleyPRE2006}. Depending on the
properties of the network and on the physical meaning of the weights
assigned to the links, we may find a wide variety of applications. For
example, for directed lattices and bond-weights representing the local
energy we get the problem of directed polymers in random media (DPRM)
\cite{Halpin_95,HansemPRL93}.\\

We can define the ball $B(t;\mathbf{x}_0)$ around a fixed node
$\mathbf{x}_0$ for time $t\geq 0$,

\begin{equation}
  B(t;\mathbf{x}_0)=\{\mathbf{x}\in\Z^2\: :\:
  T(\mathbf{x}_0,\mathbf{x})\leq t\}.
\end{equation}
Its boundary, $\partial B(t;\mathbf{x}_0)$, will be termed the {\em
  isochrone} corresponding to time $t$. Balls and isochrones can be
obtained using e.g.\ Dijkstra's algorithm \cite{Cormen}. Along this
work all the balls will be centered at the origin of coordinates
$\mathbf{x}_0=\mathbf{0}$, so we will write simply $B(t)$.

\medskip

\begin{figure}
\begin{center}
  \includegraphics[width=4.25cm]{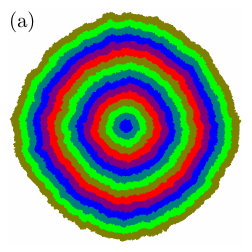}
  \hfill
  \includegraphics[width=4.25cm]{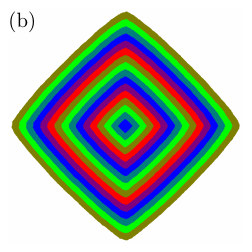}
\caption{Random balls $B(t)$ for an FPP system on a square lattice,
  centered on the origin of coordinates for different times. (a) Balls
  obtained for a uniform link-time distribution using $\mu=5$ and
  $\text{CV}=0.57$. (b) Balls obtained for a uniform link-time
  distribution using $\mu=5$ and $\text{CV}=0.11$. Colors are changed
  every $\Delta t=100$, and the total lattice size is $500\times 500$
  for panel (a) and $250\times 250$ for panel (b).}
\label{fig1_intro}
\end{center}
\end{figure}

Following Ref. \cite{Cordoba_18}, we control the strength of the
disorder througth the coefficient of variation of the distribution,
$\text{CV}$, defined as $\text{CV}=s/\mu$. It has been shown
\cite{Villarrubia_20} that for strong disorder conditions
($\text{CV}\gg 1$), the isochrones grow initially as the clusters
obtained in bond-percolation with increasing occupation probability
$p=F(t)$. Then a crossover takes place at a certain time, which
increases monotonically with the disorder strength, from which on the
isochrones evolve towards the asymptotic circular shape with KPZ
statistics. In this work we will focus on the dynamics and geometry of
the isochrones in the weak disorder regime, i.e., when $\text{CV}<1$
\cite{Cordoba_18}. The coefficient of variation has a strong effect on
the shape of the isochrones, as it is illustrated in Fig.
\ref{fig1_intro}, where we can see circular isochrones associated to
$\text{CV}=0.57$, in panel (a), and diamond-like shapes associated to
$\text{CV}=0.11$, in panel (b).

In order to characterize the isochrones, we have performed numerical
experiments on a $(2L+1)\times (2L+1)$ square lattice with $L=1200$.
In this work, we will employ two different link-time distributions.
First of all, a uniform distribution on an interval $[\tau_0,\tau_1]$,
for which the maximal attainable value of CV is $1/\sqrt{3}\approx
0.58$, since $\tau_0\geq 0$ necessarily. Furthermore, we have also
employed a Weibull distribution, given by the probability density
function

\begin{equation}
  f(\tau)=\frac{k}{\lambda}\(\frac{\tau}{\lambda}\)^{k-1}
  \exp\(-(\tau/\lambda)^k\),
\end{equation}
with shape parameter $k>1$, which is only defined for positive $\tau$,
and allows any positive value for CV. Thus, we employ a bounded and an
unbounded distribution, both of them fulfill the conditions for the
limit shape theorem, i.e. all the moments exist and are finite, and
$F(0)=0$ \cite{Cox_81,Kesten_86}.

In our simulations we fix $\mu=5$ and choose different values of
$\text{CV}$ in order to survey the different possible limit shapes.
Thus, we use the notation $U(\text{CV})$ and $Wei(\text{CV})$
respectively for the uniform and Weibull distributions with parameter
$\text{CV}$. In all the simulations discussed in this text we employ
$N_s=10^4$ different noise realizations. Measurements are performed at
$N_t=100$ logarithmically distributed times, ranging from
$t_{\text{min}}=10\mu=50$ to $t_{\text{max}}$ chosen so that the
average radius of the isochrone reaches $3L/4=900$.

%%%%%%%%%%%%%%%%%%%%%%%%%%%%%%%%%%%%%%%%%%%%%%%%%%%%%%%%%%%%%%%%%%%

\section{Shape analysis}
\label{sec:shape}

\subsection{Limit shape and instantaneous average shape}

The shape theorem ensures that the growth rates of the FPP isochrones
along any fixed direction converge towards a limiting function,
$v(\theta)$ \cite{Cox_81,Kesten_86,Damron18}. In other terms, if we
scale down the different isochrones, $t^{-1}\pl B(t)$, we will notice
that, with probability one, they are contained in a deterministic,
convex and compact set $\mathscr{B}$, which must be invariant under
reflections around the axes of the underlying lattice. Under very
general conditions, this limit shape is determined by $v(\theta)$. For
large values of CV, this limit shape will be close to a circumference,
while for small CV the average isochrone approaches a diamond shape
\cite{Cordoba_18}, given by

\beq
\{(x,y)\in\mathbb{R}^2\: :\: |x|+|y|=K\},
\label{eq:diamond}
\eeq
where $K$ is a constant ensuring that the average radius is 1,

\beq
K={\pi\over \sqrt{2}\log(2^{3/2}+3)}\approx 1.26.
\eeq

\medskip

Let us parametrize our isochrone as a polar curve for each time,
$r(\theta,t)$. The {\em mean circumference} at time $t$ will be
centered at the origin, with radius given by

\begin{equation}
  R(t)\equiv \left\langle \overline{r(\theta,t)} \right\rangle,
\end{equation}
where we denote spatial averages over $\theta$ with an overbar, and
averages over noise realizations with angular brackets. Let us then
define an {\em instantaneous average shape} (IAS) as

\begin{equation}
  R(\theta,t)\equiv \< r(\theta,t) \>,
\end{equation}
where the average is taken over all noise realizations. Notice that
this instantaneous average shape need not coincide in general with the
scaled version of the limit shape, $R(\theta,t)\neq t\,v(\theta)$. In
fact, we can define a {\em scaled IAS},

\begin{equation}
  \rho(\theta,t)\equiv \frac{R(\theta,t)}{R(t)},
\end{equation}
that will approach the limit shape asymptotically,

\begin{equation}
  \lim_{t\to\infty}  \rho(\theta,t) =C\, v(\theta),
  \label{eq:limitshape}
\end{equation}
where $C$ is a constant.

\subsection{Characterization of the instantaneous average shape}

We have evaluated the IAS $R(\theta,t)$ for different times and
disorder distributions for the link-times. For each noise realization
and time we obtain a discretized interface, whose angular resolution
$\Delta\theta$ decreases with time, because the average radius grows
with time. Indeed, we have determined our angular resolution
dynamically by imposing that the number of points in each interval
$\Delta\theta$ must remain between one and two throughout the
simulation.

\begin{figure}
\begin{center}
\includegraphics[width=8cm]{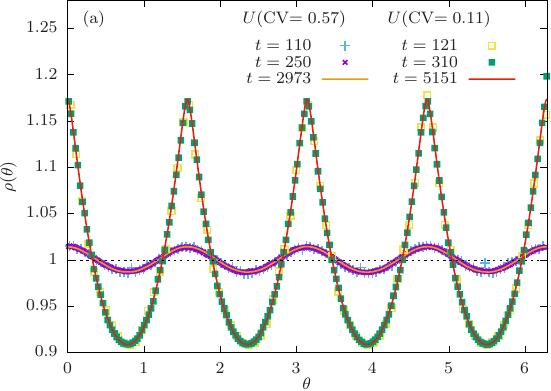}
\vskip 3mm
\includegraphics[width=8cm]{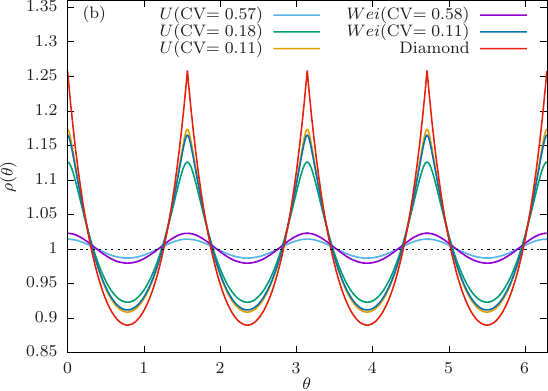}
\caption{(a) Time evolution of the scaled IAS, $\rho(\theta,t)$, for
  two uniform link-time distributions, using $\text{CV}=0.57$ and
  $\text{CV}=0.11$. (b) Limit shapes for different link-time distributions,
  all of them with expected value $\mu=5$. The dashed line represents
  the circumference, and the perfect diamond shape, given by Eq.
  \eqref{eq:diamond}, is shown with a red continuous line. }
\label{fig2}
\end{center}
\end{figure}

Figure \ref{fig2} shows the scaled IAS, $\rho(\theta,t)$, for
different times and disorder distributions as a function of the angle
$\theta$. Figure \ref{fig2} (a) shows the time evolution of the scaled
IAS for the uniform link-time distributions using $\text{CV}=0.57$ and
$\text{CV}=0.11$. Notice that the scaled IAS converges very fast to a
limit shape, as expected in Eq. \eqref{eq:limitshape}. Figure
\ref{fig2} (b) shows the limit shape obtained numerically for a
variety of uniform and Weibull distributions. As a reference, we also
show the circumference, given by the horizontal dashed line, and the
diamond, Eq. \eqref{eq:diamond}, shown as a red continuous line. We
should stress that the mean value of the link-time distribution $\mu$
does not affect the limit shape.

\begin{figure}
\includegraphics[width=8cm]{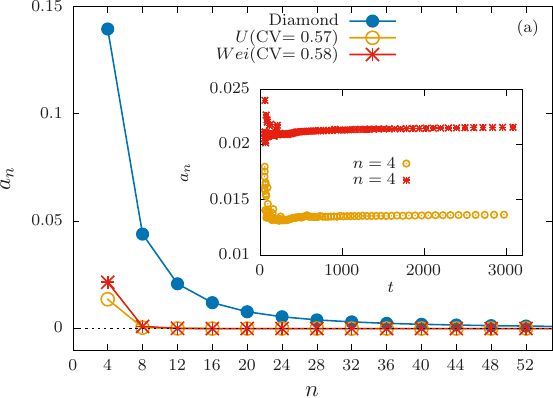}
\vskip 3mm
\includegraphics[width=8cm]{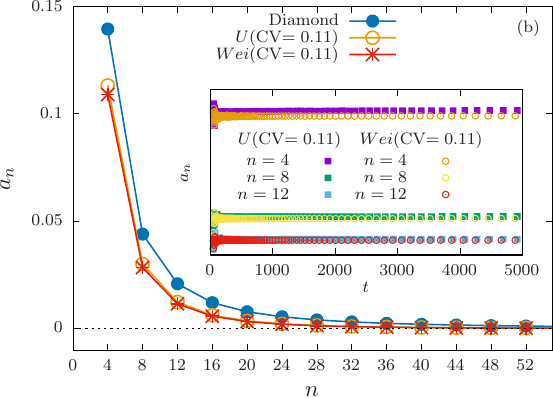}
\caption{Fourier decomposition of the limit shape, $a_n$ for the
  uniform and Weibull link-time distributions, with $\mu=5$, employing
  $\text{CV}=0.57$ in panel (a) and $\text{CV}=0.11$ in panel (b).
  Insets: Time-evolution of some selected Fourier coefficients of the
  scaled IAS, $a_n(t)$, using both $\text{CV}=0.57$ and
  $\text{CV}=0.11$ for the uniform and Weibull link-time
  distributions. }
\label{fig3}
\end{figure}

We can express the scaled IAS at time $t$, $\rho(\theta,t)$, as a Fourier
series, 

\begin{equation}
  \rho(\theta,t)=a_0+
  \sum_{n=1}^\infty \left[a_n(t)\cos\(n\theta\)
    +b_n(t)\sin\(n\theta\)\right].
\end{equation}
Let us stress that $a_0=1$ in all cases for all times, and $b_n(t)=0$
for all $n$ due to reflection symmetry around each axis. Furthermore,
$\pi/2$ rotational symmetry dictates that the only non-zero values for
$a_n(t)$ are those with $n=4k$ and $k\in\N$. Figure \ref{fig3} shows
the Fourier components for the limit shape, $a_n$, obtained for the
uniform and Weibull link-time distributions using both
$\text{CV}=0.57$ (a) and $\text{CV}=0.11$ (b). Notice that for
$\text{CV}=0.57$ only a few $a_n$ take non-zero values, while for
$\text{CV}=0.11$ we observe a behavior similar to the diamond shape,
which is also shown for comparison. Figure \ref{fig3} (insets) shows
the time evolution of different Fourier components, $a_n(t)$, obtained
for the uniform and Weibull link-time distribution using
$\text{CV}=0.57$ and $\text{CV}=0.11$, where we can observe their fast
convergence towards their limit shape values, that we will denote as
$a_n$.

%%%%%%%%%%%%%%%%%%%%%%%%%%%%%%%%%%%%%%%%%%%%%%%%%%%%%%%%%%%%%%%%%%%%%%%%%

\section{Roughness and correlation length}
\label{sec:timebehavior}

\subsection{Roughness}

\begin{figure}
\includegraphics[width=8cm]{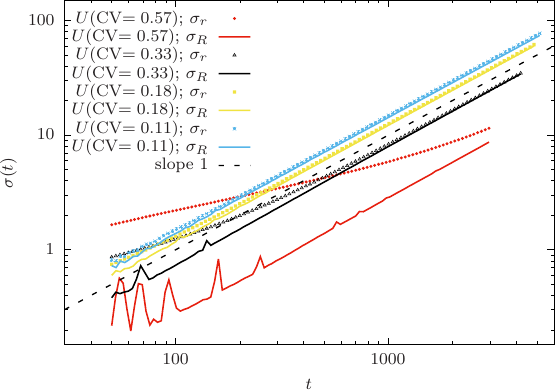}
\caption{Deviation of the interfacial radii of the isochrones,
  $\sigma_r$ (symbols) and of the IAS radii,
  $\sigma_R$ (solid lines) for several uniform
  distributions. The dashed line shows a power law behavior with
  exponent $\beta=1$.} 
\label{fig4}
\end{figure}

Let us now consider the roughness and the correlation lengths for the
isochrones. In naive terms, we may define the isochrone roughness
corresponding to time $t$ as the deviation of all values $r(\theta,t)$
around their average value, $R(t)$ \cite{Barabasi},

\begin{equation}
  \sigma^2_r(t) \equiv
  \<\overline{\(r(\theta,t)-R(t)\)^2}\>,
\label{eq:naive_roughness}
\end{equation}
Indeed, this definition is not appropriate because the average shape
is not circular. Thus, the roughness should be measured with respect
to a suitable average shape. In fact, Fig. \ref{fig4} shows the time
evolution of this naive roughness (solid lines), which grows with time
as $t^\beta$ with $\beta=1$, very different from the expected
$\beta=1/3$ KPZ value \cite{Cordoba_18}. The reason for this growth is
that most of the radial deviation can be associated to the deviations
of the IAS with respect to the mean circumference. Indeed, Fig.
\ref{fig4} shows also the deviation of the IAS radii, given by

\begin{equation}
  \sigma^2_R(t) \equiv \< \overline{\(R(\theta,t) - R(t)\)^2} \>,
  \label{eq:ias_roughness}
\end{equation}
which presents a similar scaling for small $\text{CV}$. Even for higher values
of $\text{CV}$ we can observe that both deviations approach asymptotically
for long times. Therefore, we can explain the $\beta=1$ scaling: we
are counting as roughness what is simply the form of the average
isochrone, which is due to the lattice.

\medskip

\begin{figure}
\begin{center}
\includegraphics[width=8cm]{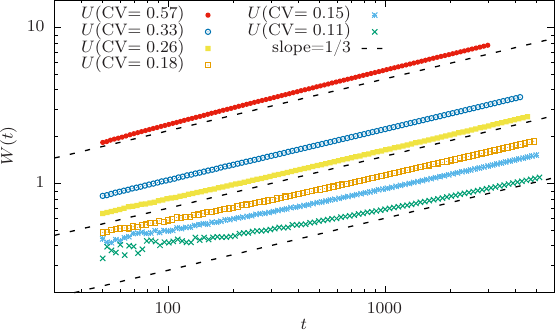}
\caption{Time evolution of the roughness $W(t)$, defined with respect
  to the IAS, as shown in Eq. \eqref{eq:true_W}, for uniform
  distributions. Dashed lines indicate power-law behavior with
  exponent $1/3$.}
\label{fig5}
\end{center}
\end{figure}

Let us provide the correct definition of the interface roughness,
which is the root-mean-square deviation between the radii of the
interfaces and the IAS,

\beq
W^2(t) \equiv  \< \overline{\(r(\theta,t)-R(\theta,t)\)^2 }\> .
\label{eq:true_W}
\eeq
Figure \ref{fig5} shows the time evolution of this roughness $W(t)$ for
different noise distributions. As expected, this magnitude presents a
scaling behavior associated to the KPZ universality class, $W(t)\sim
t^\beta$ with $\beta=1/3$ in all cases. We should remark that for higher
values of $\text{CV}$ the exponent $1/3$ is rapidly attained, whereas the
preasymptotic regime becomes longer for lower values of $\text{CV}$,
%. This parameter $d_c$, 
as expected from previous works \cite{Cordoba_18}.

\subsection{Angular-resolved roughness}

\begin{figure}
\includegraphics[width=8cm]{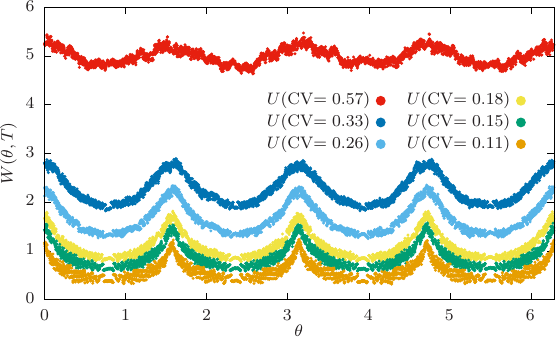}
\caption{Angular-resolved roughness, $W(\theta,t)$, for time
  $t_{\text{max}}$, using different uniform link-time distributions.}
\label{fig:angular_resolved}
\end{figure}

Our definition of the roughness, given in Eq. \eqref{eq:true_W}, can
be considered as an angular average of a certain angular-resolved
roughness, which can be defined as

\beq
W^2(\theta,t)\equiv \< \(r(\theta,t)-R(\theta,t)\)^2 \>,
\label{eq:wtheta}
\eeq
such that

\beq
W(t)=\overline{W(\theta,t)}.
\eeq
Indeed, in Fig. \ref{fig:angular_resolved} we can observe the
angular-resolved roughness $W(\theta,t)$ obtained at time
$t_{\text{max}}$ using several uniform link-time distributions. We
notice that this roughness presents a strong anisotropy for small
values of $\text{CV}$, as expected. Indeed, the roughness is always
larger near the lattice axes.

\subsection{Correlation length}

On the other hand, the scaling of the correlation length with time
allows us to obtain the dynamic exponent $z$, from the evolution of
the correlation length, $\xi \sim t^{1/z}$. We will estimate its value
using the technique developed in Ref.~\cite{Santalla_17}, which is
based on the notion of {\em patches}. A patch is defined as a section
of the isochrone such that all its points are either above or below
the IAS. The patch length is then defined as the projection of this
subset of the isochrone onto the average circumference. Let $n$ be the
number of patches of a given isochrone, whose lengths are given by
$\{\ell_i\}_{i=1}^n$. In order to estimate the correlation length we
choose randomly a point over the average shape and find the expected
value of the length of its associated patch. In other words, we select
patch $i$ with probability $\ell_i/\sum_i \ell_i$, and we can estimate

\begin{equation}
  \xi\equiv \<\frac{\sum_i \ell_i^2}{\sum_i \ell_i}\>,
\label{correlacion}
\end{equation}
which is expected to grow as $\xi(t)\sim t^{1/z}$.

\begin{figure}
\begin{center}
\includegraphics[width=8cm]{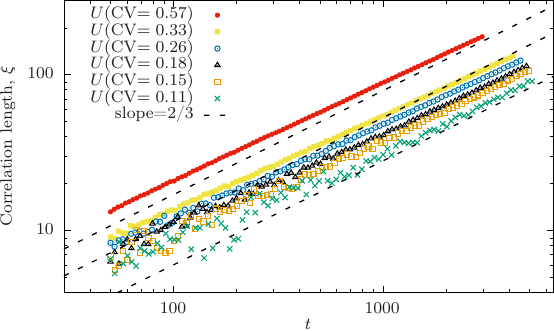}
\caption{Growth of the correlation length for several uniform
  distributions. Dashed lines show a power-law behavior with exponent
  $2/3$. }
\label{fig6}
\end{center}
\end{figure}

We have evaluated the behavior of $\xi(t)$ for all the considered
distributions of disorder, and the results for several uniform
distributions are displayed in Fig. \ref{fig6}. We obtain in all cases
a power-law with an exponent very close to $2/3$. These results are in
agreement with a value for the dynamic exponent of $z=3/2$, associated
to the KPZ universality class.

%%%%%%%%%%%%%%%%%%%%%%%%%%%%%%%%%%%%%%%%%%%%%%%%%%%%%%%%%%%%%%%%%%%%%

\section{Radial fluctuations}
\label{sec:fluct}

As it is discussed in the introduction, the height fluctuations in
1+1D KPZ systems follow the Tracy-Widom (TW) probability distribution
\cite{Takeuchi_11}, which somehow plays a similar role to that of the
Gaussian distribution in the central limit theorem. The TW
distribution comes in different flavors, and the Gaussian orthogonal
ensemble (TW-GOE) is typically associated to flat interfaces, while
the Gaussian unitary ensemble (TW-GUE) is typically connected to
circular interfaces.

In a model with rotational symmetry, we can consider all the radii of
the different interfaces, labeled by their time, $\{r_i(t)\}$. Then,
we fit the average radius as a function of time to a form
$\<r_i(t)\>\approx r_0+vt$, choosing the best possible values of $r_0$
and $v$. Then, we fit the time dependence of their deviations to a
form $\sigma[r_i(t)] \approx \Gamma t^\beta$. Finally, we define a
random variable $\chi_i$ implicitly through the expression,

\beq
r_i=r_0 + vt + \Gamma t^\beta \chi_i.
\eeq
If the interface follows KPZ scaling, we expect the random variable
$\chi$ to present a stationary probability distribution that will
approach the TW-GUE distribution, rescaled to have zero average and
unit deviation \cite{Santalla_15,Santalla_17}.

The same TW-GUE distribution has been found in FPP models, for
example, in the times of arrival along the axis and diagonal
directions \cite{Cordoba_18}. Yet, KPZ scaling suggests that it should
also determine the radial fluctuations of the isochrones, and in this
section we will show that this is indeed the case, provided that these
radial fluctuations are appropriately scaled.

\medskip

Let us choose a point $\mathbf{x_i}=(r_i,\theta_i)$ along the
interface, and let $R(\theta,t)$ be the radius of the IAS along the
same direction for that time. Then, we can define

\begin{equation}
  r=R(\theta,t)+ \Gamma(\theta) t^{1/3}\chi,
  \label{eq:radialfluct}
\end{equation}
where $\chi$ should be a stationary random variable following the
(rescaled) TW-GUE distribution and $\Gamma(\theta)$ is chosen so that
the angular-resolved roughness, defined in Eq. \eqref{eq:wtheta},
behaves as

\beq
W(\theta,t) \approx \Gamma(\theta) t^\beta.
\eeq
Thus, in our case we must employ {\em two labels} for the radial data,
i.e. the time and the angle $\theta$.

\begin{figure}
\includegraphics[width=8cm]{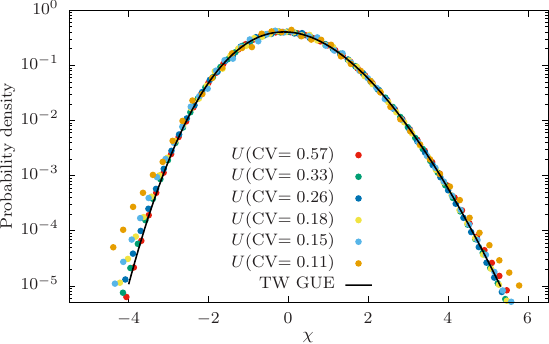}
\caption{Histograms for the rescaled radial fluctuations, following
  Eq. \eqref{eq:radialfluct} for several uniform distributions.}
\label{fig:tw}
\end{figure}

Numerically, we proceed as follows. We select a certain angular width
$\Delta\theta\approx 10^{-3}$, and bin our radial data according to time and
angle. For each bin, we subtract the expected average, which
corresponds to the radius of the IAS, and divide by their deviation,
which corresponds to the associated angular-resolved roughness,

\beq
\chi_i = {r_i - R(\theta,t) \over W(\theta,t) }.
\eeq
We then obtain the histograms for these values in Fig. \ref{fig:tw},
which corresponds to the TW-GUE distribution as expected. The
histograms are computed using all available times for which the
condition $W(t)\thicksim t^{\beta}$ holds.

\begin{figure}
\begin{center}
\includegraphics[width=8cm]{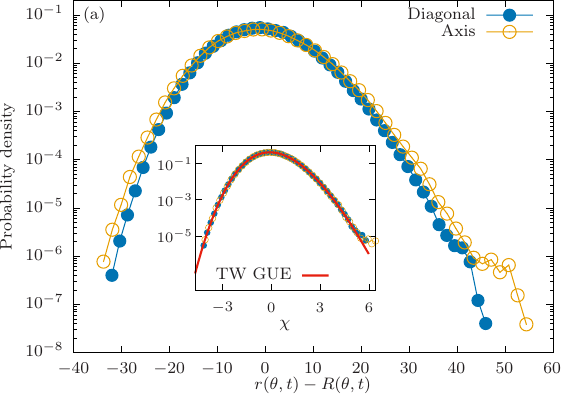}
\vskip 3mm
\includegraphics[width=8cm]{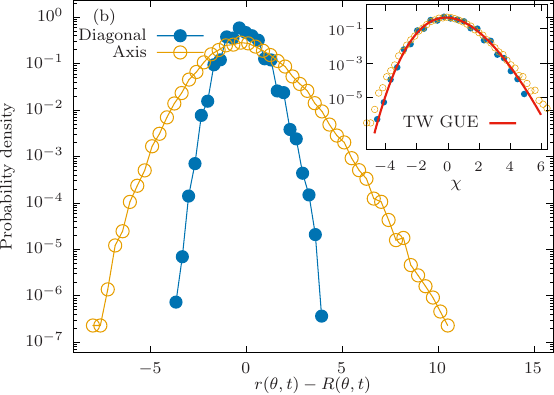}
\caption{Histogram of the radial fluctuations,
  $r(\theta,t)-R(\theta,t)$, without proper rescaling with the
  deviation, performed for the axial and diagonal directions with an
  angular window of $\Theta=1/3$ radians, using (a)
  $U(\text{CV}=0.57)$, (b) $U(\text{CV}=0.11)$, for time
  $t_{\text{max}}$. Insets: Histograms of the rescaled radial
  fluctuations for the same directions.}
\label{fig9}
\end{center}
\end{figure}

Figure \ref{fig9} shows the histograms for the radial fluctuations,
$r(\theta,T)-R(\theta,T)$, without rescaling with the corresponding
deviation, along the axis and diagonal directions for different values
of $\text{CV}$. The measurement time is $t_{\text{max}}$, and in order
to obtain enough data we employ angular windows of width $\Theta=1/3$
radians along both directions. Panel Fig. \ref{fig9} (a) shows the
histograms for $\text{CV}=0.57$, and we can see that the distributions
for both directions are rather similar, both corresponding to the
TW-GUE distribution as expected. The inset panel shows how they
coincide when the fluctuations are rescaled to have unit variance.
Panel Fig. \ref{fig9} (b), on the other hand, shows the histograms for
$\text{CV}=0.11$, where we can observe that the fluctuations along the
axis and the diagonal directions are very different, while they
coincide when correctly rescaled, as shown in the inset. A naive
averaging of the fluctuations along different directions, without
proper rescaling with the appropriate deviations, would lead to an
average histogram which departs enormously from the TW-GUE
distribution, showing indeed a large value for the kurtosis.

%%%%%%%%%%%%%%%%%%%%%%%%%%%%%%%%%%%%%%%%%%%%%%%%%%%%%%%%%%%%%%%%%%%

\section{Morphological analysis}
\label{sec:morphology}

Let us consider the roughness exponent, $\alpha$, which characterizes
the stationary regime attained when the correlation length reaches the
system size $L$. According to the Family-Vicsek dynamic scaling Ansatz
\cite{Barabasi}, we have

\begin{equation}
W(L,t)=t^\beta \;g\(L/t^{1/z}\),
\end{equation}
where the scaling function $g(u)$ has the general form

\begin{equation}
g(u)=\left\{
\begin{aligned}
\text{const.}\qquad&\text{if } u\gg 1, \\
u^{\alpha}\qquad&\text{if } u\ll 1. \\
\end{aligned}
\right. 
\label{eq:global_fv}
\end{equation}
This expression can not be applied to radially growing systems in
which the system size increases linearly with time, because in those cases
the stationary regime in never attained \cite{Bru03,Santalla_14}. Yet,
the power-law behaviors of the roughness $W(t) \sim t^{\beta}$ and the
correlation length $\xi(t)\sim t^{1/z}$, along with the Galilean
scaling relation for KPZ, $\alpha+z=2$, all suggest $\alpha=1/2$ for
our system, but we should check this value independently. 

In order to characterize the morphological properties of the
isochrones, we define the scale-resolved roughness, $w(l,t)$, as the
average roughness for windows of size $l$ measured on the average
circumference of radius $R(t)$ \cite{RamascoPRL2000},

\beq
w^2(l,t)\equiv \< \left[ \(r(\theta,t) - R(t)\)^2 \right]_l \>,
\label{eq:local_roughness}
\eeq
where $\left[\cdots\right]_l$ denotes the average over linear windows
of size $l$, whose location does not depend on $\theta$. We expect the
following scaling ansatz, similar to Eq.~\eqref{eq:global_fv},

\beq
w(l,t)=t^{\beta}g_{\text{loc}}(l/t^{1/z}),
\eeq
where now $g_{\text{loc}}(u)$ behaves as $u^{\alpha_l}$ if $u\ll 1$ and as a
constant for $u\gg 1$, where $\alpha_l$ is termed the local roughness
exponent. Thus, for a fixed value $l$ we expect the local roughness to
grow as $t^\beta$ up to a saturation time, when the correlation length
reaches $l$. From this moment on the window roughness saturates at a
value that scales as $l^{\alpha_l}$.

\medskip

\begin{figure}
\begin{center}
\includegraphics[width=8cm]{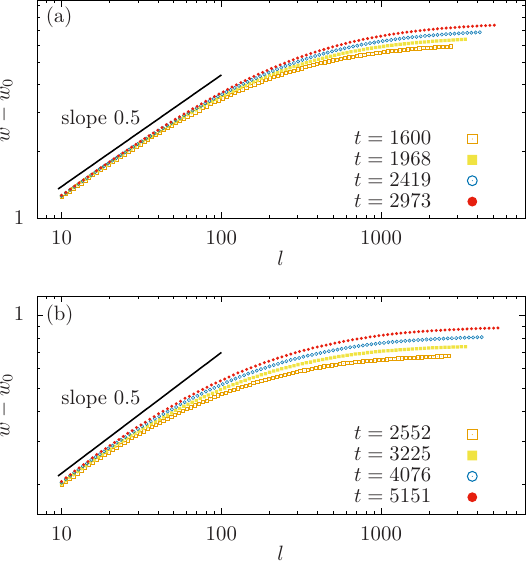}
\caption{Scale-resolved roughness, subtracting the intrinsic
  roughness, $w(l,t)-w_0$, as a function of $l$ for disorder
  distributions $U(\text{CV}=0.57)$ (a) and $U(\text{CV}=0.11)$ (b).}
\label{fig:alpha}
\end{center}
\end{figure}

We should adapt our measurements of the local roughness, Eq.
\eqref{eq:local_roughness}, to our anisotropic case,

\beq
w(l,t) = \< \left[ \(r(\theta,t) - R(\theta,t)\)^2 \right]_l \>.
\label{eq:local_roughness_2}
\eeq
The results of our numerical simulations are shown in Fig.
\ref{fig:alpha}, where the scale-resolved roughness is plotted for
different times, using uniform link-distributions with
$\text{CV}=0.57$ and $\text{CV}=0.11$. The proper small-lengths
scaling is obtained assuming that the system presents certain {\em
  intrinsic roughness},

\beq
w(l,t) \approx w_0 + A l^\alpha,
\eeq
where $\alpha=1/2$ in all the cases, $w_0\approx 0.17$ for
$\text{CV}=0.57$, and $w_0\approx 0.2$ for $\text{CV}=0.11$. This
intrinsic roughness can mask the correct scaling if it is not properly
taken into account. We conjecture that its physical origin is related
to the uncertainty in the measurement of the IAS.

%%%%%%%%%%%%%%%%%%%%%%%%%%%%%%%%%%%%%%%%%%%%%%%%%%%%%%%%%%%%%%%%%%%%

\section{Conclusions and further work}
\label{sec:conclusions}

In this article we have characterized the statistical properties of
the isochrones of the first-passage percolation (FPP) problem on a
square lattice, showing that they correspond to the KPZ universality
class. The main difficulty lies in the fact that the average isochrone
deviates substantially from the circumference when the coefficient of
variation (CV) is small, due to the strong anisotropy of the system.
In order to reveal the hidden KPZ scaling, we have defined the
instantaneous average shape (IAS) for each noise level and time, and
characterized them using their Fourier representation. Indeed, for
$\text{CV}\approx 1$ they approach a circumference, while for
$\text{CV}\ll 1$ they approach a {\em diamond}. We also define an
angular-resolved roughness, which depends on the direction of growth,
showing a similar behavior. Indeed, the fluctuations are always higher
along the axis than along the diagonal, with the anisotropy again
growing for lower values of $\text{CV}$.

Once the radial fluctuations are measured with respect to the IAS, and
their deviations are scaled with an angular-dependent factor, all the
hallmarks of the 1+1D KPZ class appear clearly: the roughness grows as
$W(t) \sim t^\beta$ with the growth exponent $\beta=1/3$, and the
correlation length grows as $\xi(t)\sim t^{1/z}$ with the dynamic
exponent $z=3/2$. Moreover, if the radial fluctuations are scaled with
an angular-dependent factor, they are shown to follow the expected
TW-GUE distribution. The local roughness can be shown to scale as
$w(l)\sim l^{\alpha_l}$ for $l\ll\xi(t)$, with $\alpha_l=1/2$, if we
subtract previously an intrinsic roughness $w_0$ which depends weakly
on the link-time distribution, and which is probably related to the
uncertainty in our estimation of the IAS.

Our procedure to subtract the IAS will be of interest in order to
analyze other systems, both continuous or discrete, which present
non-circular characteristic shapes around which the fluctuations
should be measured. The first such application should be to growing
interfaces defined on a lattice, when the lattice effects are
suspected to spoil the scaling analysis and the characterization of
the universality class \cite{Ferreira.06,Bonn.09,Marcos.22}.
Physically, lattice effects may be induced by e.g. crystallographic
directions, as in epitaxial growth of thin solid films
\cite{Jensen.99,Misbah.10}. Furthermore, the subtraction of the
average shape can be useful in cases of morhopological instabilities,
where a non-trivial average shape can be established at short times,
which may in turn have an effect in the determination of the universal
properties of the fluctuations \cite{Castro.12,Santalla.18}.

%%%%%%%%%%%%%%%%%%%%%%%%%%%%%%%%%%%%%%%%%%%%%%%%%%%%%%%%%%%%%%%%%%%%%%%%%%

\begin{acknowledgments}
This work was partially supported by Ministerio de Ciencia e
Innovaci\'on (Spain), Agencia Estatal de Investigaci\'on (AEI, Spain,
10.13039/501100011033), and European Regional Development Fund (ERDF,
A way of making Europe) through Grants Nos. PID2019-105182GB-I00 and
PID2021-123969NB-I00, and by Comunidad de Madrid (Spain) under the
Multiannual Agreement with UC3M in the line of Excellence of
University Professors (EPUC3M14 and EPUC3M23), in the context of the V
Plan Regional de Investigaci\'on Cient\'{\i}fica e Innovaci\'on
Tecnol\'ogica (PRICIT). We acknowledge the computational resources and
assistance provided by the Centro de Computaci\'on de Alto Rendimiento
CCAR-UNED. I.A.D.\ acknowledges funding from UNED through an FPI
scholarship.
\end{acknowledgments}

%%%%%%%%%%%%%%%%%%%%%%%%%%%%%%%%%%%%%%%%%%%%%%%%%%%%%%%%%%%%%%%%%%%%%%%%%%

\end{document}